\RequirePackage{fix-cm}
\RequirePackage{rotating}
\documentclass[smallextended]{svjour3}       % onecolumn (second format)
\smartqed  % flush right qed marks, e.g. at end of proof
\usepackage{graphicx}
\usepackage{rotating}
\usepackage{lineno,hyperref}
\usepackage{amsmath}
\usepackage{booktabs}
\usepackage{multirow}
\usepackage{multicol}
\usepackage{graphicx}
\usepackage{pdfpages}
\usepackage{subcaption}
\usepackage[newcommands]{ragged2e}

\begin{document}
	
	\title{Predictive Factors of Kinematics in Traumatic Brain Injury from Head Impacts Based on Statistical Interpretation}
	
	% \subtitle{}

	\author{Xianghao Zhan \and Yiheng Li \and Yuzhe Liu \and August G. Domel \and Hossein Vahid Alizadeh \and Zhou Zhou \and Nicholas J. Cecchi \and Samuel J. Raymond \and Stephen Tiernan \and Jesse Ruan \and Saeed Barbat \and Olivier Gevaert \and Michael M. Zeineh \and Gerald A. Grant \and David B. Camarillo}

	\institute{Xianghao Zhan, Yuzhe Liu, August G. Domel, Hossein Vahid Alizadeh, Zhou Zhou, Nicholas J. Cecchi, Samuel J. Raymond , David B. Camarillo \at
		Department of Bioengineering, Stanford University, 94305, CA, USA. \\
		\email{xzhan96@stanford.edu, yuzheliu@stanford.edu, augustgdomel@gmail.com, hva@stanford.edu, zhouz@stanford.edu, ncecchi@stanford.edu, sjray@stanford.edu, dcamarillo@stanford.edu}           
		\and
		Yiheng Li, Olivier Gevaert \at
		Department of Biomedical Data Science, Stanford University, 94305, CA, USA. \\ 
		\email{yyhhli@stanford.edu, ogevaert@stanford.edu}
		\and
		Stephen Tiernan \at
		Technological University Dublin, Dublin, Ireland.\\
		\email{stephen.tiernan@tudublin.ie}
		\and
		Jesse Ruan, Saeed Barbat \at
		Ford Motor Company, 3001 Miller Rd, Dearborn, MI 48120, USA.
		\email{jruan@ford.com, sbarbat@ford.com}
		\and
		Michael M. Zeineh \at
		Department of Radiology, Stanford University, 94305, CA, USA.
		\email{mzeineh@stanford.edu}
		\and
		Gerald A. Grant \at
		Department of Neurosurgery, Stanford University, 94305, CA, USA.
		\email{ggrant2@stanford.edu}
		\and
		Xianghao Zhan and Yiheng Li contributed equally to this work.
	}
	
	\date{Received: date / Accepted: date}
	
	\maketitle
	
	\begin{abstract}
		Brain tissue deformation resulting from head impacts is primarily caused by rotation and can lead to traumatic brain injury. To quantify brain injury risk based on measurements of kinematics on the head, finite element (FE) models and various brain injury criteria based on different factors of these kinematics have been developed, but the contribution of different kinematic factors has not been comprehensively analyzed across different types of head impacts in a data-driven manner. To better design brain injury criteria, the predictive power of rotational kinematics factors, which are different in 1) the derivative order (angular velocity, angular acceleration, angular jerk), 2) the direction and 3) the power (e.g., square-rooted, squared, cubic) of the angular velocity, were analyzed based on different datasets including laboratory impacts, American football, mixed martial arts (MMA), NHTSA automobile crashworthiness tests and NASCAR crash events. Ordinary least squares regressions were built from kinematics factors to the 95\% maximum principal strain (MPS95), and we compared zero-order correlation coefficients, structure coefficients, commonality analysis, and dominance analysis. The angular acceleration, the magnitude and the first power factors showed the highest predictive power for the majority of impacts including laboratory impacts, American football impacts, with few exceptions (angular velocity for MMA and NASCAR impacts). The predictive power of rotational kinematics about three directions (x: posterior-to-anterior, y: left-to-right, z: superior-to-inferior) of kinematics varied with different sports and types of head impacts.
	\end{abstract}
	
	\section{Introduction}
	Traumatic brain injury (TBI), a primary cause of death and disability, poses a global health threat and affects over 1.7 million children and adults in the United States and over 55 million people worldwide \cite{Spencer16}. TBI may arise from falls, vehicle accidents, and popular contact sports such as American football and mixed martial arts (MMA) \cite{Montenigro17,OKeeffe19}. The prevalence of TBI and the severe consequences suffered by TBI patients calls for better monitoring of brain injury risks because early detection and intervention are essential for recovery \cite{guiza2017}. 
	
	To evaluate brain injury risk, several brain injury criteria (BIC) have been proposed based on reduced-order mechanics models \cite{Panzer1,Gabler2019,BAM} and deep learning models \cite{Shaoju,MLHM,Ji20}. However, great disparities are noted among the mathematical forms of the existing kinematic-based BIC in terms of the factors (combinations of kinematic features) \cite{HIC,BrIC,RIC}, such as: 1) the derivative order of angular velocity, e.g., the Brain Injury Criterion (BrIC) includes the peak values of angular velocity \cite{BrIC} and the Rotational Injury Criterion (RIC) includes the integral value of the magnitude of angular acceleration \cite{RIC}; 2) the different components of kinematics, e.g., BrIC includes the kinematics in three spatial directions \cite{BrIC} while RIC includes only the magnitude \cite{RIC}; 3) power of kinematics, e.g. RIC includes the integral values with a power of 2.5 \cite{RIC}. Although various factors have been included in different model designs, the contribution of each of the factors is not yet fully understood in this field, particularly when multiple datasets of various head impact types are taken into consideration and compared together.
	
	Though a complete understanding for the mechanobiological cascade from the exterior mechanical to resultant brain injury remained to be further explored, it has been suggested that strain-based metrics correlated with brain injury \cite{OKeeffe19,Bain00,Bar-Kochba16,Cater06,Donat21,Fahlstedt15,Gennarelli89,Hajiaghamemar21,Hernandez19,Kang15,McAllister12}. For example, by relating the controlled mechanical inputs to ensuing functional impairment or viability of the hippocampal cells in the \emph{in vitro} models, it is found that functional tolerance of hippocampal cells was dependent on both strain and strain rate \cite{Kang15}, while cell death on strain, not on strain rate \cite{Cater06,Hernandez19} leveraged an FE model of human head to predict the intracranial response during the head impacts in football game and found that the model-estimated fiber strain (i.e., normal deformation along the fiber tracts) correlated with the callosal abnormalities revealed by diffusion tensor imaging of the athletes. Recently, \cite{Hajiaghamemar21} employed an FE model of pig brain to simulate a set of pig experiments with the experimental loading and resultant pathological distribution well documented. A spatial correlation was noted between the large fiber strain obtained from the FE model and the acute axonal injury sustained by the experimental pigs after rapid head rotations. Therefore, brain strain, particularly maximum principal strain, is regarded as a key parameter indicating brain injury risk for TBI research \cite{Panzer1,Gabler2019,Shaoju,InertialF}.
	
	Previous studies have evaluated the BIC by comparing them with the 95th percentile of maximum principal strain (MPS95) \cite{Panzer1,Gabler2019,Shaoju,MLHM,Ji20,BIC} and many recently developed black-box deep-learning-based models can accurately estimate the MPS95 \cite{Shaoju,MLHM}. Although based on a wide range of kinematic features, the deep learning models are accurate, they are black-box, which cannot explicitly show the contribution of different kinematic factors. Based on the finite element (FE) modeling and BIC, several studies have investigated the contribution of different kinematic factors \cite{Panzer1,con1,con2,con3,con4,con5}. For example, with a single-degree-of-freedom mechanical model for predicting strain-based brain injury responses and a dataset of idealized rotational pulses, Gabler et al. \cite{Panzer1} found that for short-duration pulses, maximum brain deformation was primarily dependent on angular velocity magnitude. While for long-duration pulses, the angular acceleration magnitude stood out. However, these studies mainly rely on single type of head impact data (e.g. football impacts, MMA impacts), and not many statistical interpretation tools other than linear correlation were used. Therefore, the predictive power of different kinematic factors in the regression of the MPS95 was not comprehensively studied and compared across a wide range of different types of head impacts (e.g. different contact sports) in a data-driven manner. To address this problem, the present study aims to statistically analyze the predictive power of the different kinematic factors in the regression of brain strain over a broad range of candidate factors in a data-driven manner.
	
	In this study, the kinematics from simulated head model impacts, American football impacts, MMA impacts, car crash impacts, and racing car impacts were used. On each dataset, four statistical interpretation methods were used: zero-order correlation coefficients, structure coefficients, commonality analysis, dominance analysis. The first two methods analyze the predictive power of individual kinematic factors and the last two methods analyze the predictive power of combinations of kinematic factors. The features used in the linear regression were derived from angular velocity, because linear acceleration has been proved to contribute less to brain strain \cite{Holbourn}, and the reference MPS was calculated by the KTH model \cite{Kleiven2007}, which is one validated model of finite element analysis (FEA).
	
	We analyzed the contributions of the kinematic factors based on three different angles: 1) the derivative order (zero-order/first-order/second-order derivative of the angular velocity $\omega(t)$: angular velocity $\omega(t)$, angular acceleration $\alpha(t)$, angular jerk $j(t)$); 2) the components in three directions in the anatomical reference frame and the magnitude; and 3) the power (square root, quadratic, cubic, etc.). The factors were listed in Table 1. For the analysis from one angle, the different factors from other angles were combined: for example, when analyzing the components, we analyzed angular velocity, angular acceleration, and angular jerk together.
	
	\begin{table}
		\centering
		\caption{The definition of the kinematic factors derived from angular velocity in this study. The factors were different in 1) the derivative order (angular velocity, angular acceleration, angular jerk); 2) the components in three directions in the anatomical reference frame and the magnitude; and 3) the power (square root, quadratic, cubic, etc.).}
		\label{tab:2}       % Give a unique label
		\begin{tabular}{cccc}
			\hline\noalign{\smallskip}
			Aspect & Kinematic factors & Further explanation & Factor number\\
			\noalign{\smallskip}\hline\noalign{\smallskip}
			Derivative order &  $\omega(t)$; $\omega^{'}(t)$; $\omega^{''}(t)$ & $\omega(t);\alpha(t);j(t)$ & 3 \\
			Components & $f_{x}(t)$; $f_{y}(t)$; $f_{z}(t)$; $|f(t)|$ & $f = \omega, \alpha, j$ & 4\\
			Power & $f^n(t)$ & $n = 0.5, 1, 2, 3, 4, 5, 6, f = \omega, \alpha, j$ & 7\\
			\hline\noalign{\smallskip}
		\end{tabular}
		% 		\tablecomments{\omega = \text{angular velocity} \\ 
		% 		               \alpha = \text{angular acceleration} \\ 
		% 		               j = \text{angular jerk} \\
		% 		               }
		\label{tab:def}
	\end{table}

	\section{Materials and Methods}
	\subsection{Data description}
	To explore the predictive factors in the prediction of MPS95 and find commonality and dissimilarities across various types of head impacts, the kinematics of head impacts from different sources were used: 2130 simulated head impacts from a validated finite element analysis (FEA) model of the Hybrid III anthropomorphic test dummy headform \cite{MLHM,DummyFEA} (dataset HM). The dataset involved impacts from different locations with the velocities ranging from 2 m/s to 8 m/s. More details of the 2130 impacts can be found in \cite{MLHM}; 184 college football head impacts collected by the original version of the Stanford instrumented mouthguard \cite{camarillo2013} (dataset CF1); 118 college football head impacts collected by the updated version of the Stanford instrumented mouthguard \cite{MGValidation} (dataset CF2); 457 mixed martial arts (MMA) collected by the updated version of the Stanford instrumented mouthguard \cite{OKeeffe19,MMA2} (dataset MMA); 53 reconstructed head impacts by the National Football League (NFL) \cite{NFL} (dataset NFL); 48 car crash dummy head impacts from NHTSA \cite{NHTSA} (dataset NHTSA); 272 numerically reconstructed head impacts in National Association for Stock Car Auto Racing (NASCAR) by Hybrid III ATD headform (dataset NASCAR).
	
	The KTH model (Stockholm, Sweden) \cite{Kleiven2007} was used to simulate brain deformation. This model was developed in LS-DYNA (Livermore, CA, USA), including the brain, skull, scalp, meninges, falx, tentorium, subarachnoid cerebrospinal fluid (CSF), ventricles, and 11 pairs of the largest bridging veins. Responses of the head model have correlated well with experimental data of brain-skull relative motion \cite{Kleiven2006}, intracranial pressure \cite{Kleiven2006}, and brain strain \cite{zhou2018}. In this study, a total of 3262 impacts from the 7 datasets described above were simulated using the KTH head model. For each simulation, the 95th percentile maximum principal strain across all brain elements (MPS95) was extracted, which indicated the maximum deformation at the whole-brain level. \cite{Panzer1,Gabler2019}. 
	
	\subsection{Rotational kinematics features}
	To build the regression models of MPS95, various features were extracted. Because linear acceleration proved to contribute less to MPS \cite{InertialF,Holbourn}, we limited this study within the rotational kinematics and the feature extraction was done based on angular velocity. All features were peak values in time. To analyze the contribution from different rotational parameters, different kinematic components in different directions, and the non-linear relationship between kinematics and MPS95, the following factors were extracted: 1) the derivative order; 2) the components in three directions and the magnitude; 3) the power. In this study, a feature was defined as one specific value calculated based on the rotational kinematics with definite derivative order, component, and power; a factor was a group of features in each level of analysis with a shared derivative order/component and magnitude/power. For instance, the time-peak value of the angular velocity in the x-axis direction was a feature, and all the time-peak values of the angular velocity (including three directions and the magnitude and all different powers) formed a factor. The feature details are elaborated as follows:
	
	1) For the derivative order of angular velocity, the original angular velocity (zero-order derivative), the first-order and the second-order derivatives (respectively angular acceleration and angular jerk) were included in the analysis: $\omega(t)$, $\omega^{'}(t)$, $\omega^{''}(t)$. The numerical difference was calculated by a 5-point stencil derivative equation. This analysis was limited to these three derivative orders because they have definite physical meanings, i.e., the angular velocity, the angular acceleration, and the angular jerk. Furthermore, the numerical differentiation generally introduces noise, which makes the high-order derivative of the angular velocity inaccurate. 
	
	2) The components of rotational parameters were derived from the 3D kinematics in the anatomical reference frame (x: posterior-to-anterior, y: left-to-right, z: superior-to-inferior). The magnitude of the kinematics was also calculated and used as the fourth component. It has been shown that impacts from the side and rear directions might be more likely to lead to obvious performance decrement (OPD) in football players \cite{Bartsch2020}, which suggests that the specific directions of 3D kinematics should be included in the regression of brain strain.
	
	3) To capture the non-linear relationship between the features and MPS95, powers of $1, 0.5, 2, 3, 4, 5, 6$ of the features mentioned above were calculated and added to the feature set, respectively. Higher orders of power were ignored because there was a trend of decreasing contribution in the regression model as the order increased. The powers larger than 6 were assumed to be less predictive.
	
\begin{sidewaysfigure}
		\centering
		\includegraphics[
		page=1,
		width=0.9\linewidth,
		keepaspectratio,
		trim=0 0 0 0pt
		]{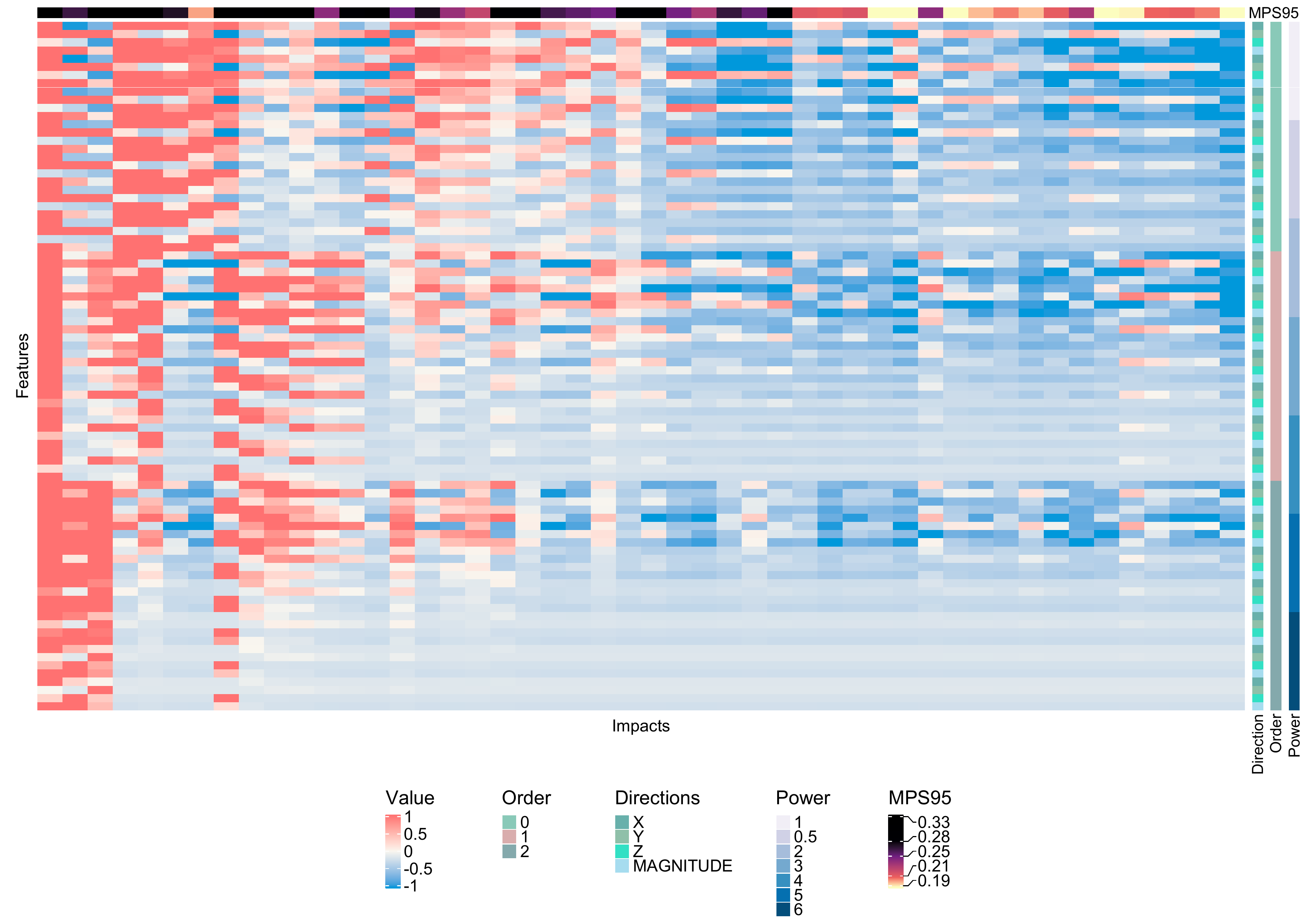}
		\caption[short caption]{Heatmap of features in dataset NHTSA. The features were standardized and the values denote how much a feature value deviates from the mean feature value. The heatmap also shows the MPS95 for each sample at the top and the different factors of the feature on the right.}
		\label{fig:heatmap}
	\end{sidewaysfigure}

	All features were based on the absolute values, and therefore, both negative peaks and positive peaks were considered. Collectively, 84 features were extracted from the kinematics in different derivative orders (3 factors), component and the magnitude (4 factors), and powers (7 factors). For each dataset, feature-wise standardization was applied to normalize different units. Additionally, standardization of predictors was also required when applying certain statistical interpretation methods \cite{UMN}, which will be discussed in Section 2.3. After feature standardization, the features and the corresponding MPS95 of each impact were visualized in the example heatmap (Fig. \ref{fig:heatmap}) that described dataset NHTSA. (Summary statistics of the kinematics and MPS95 were shown in Supplementary Figs. 1. Heatmaps of other datasets were shown in Supplementary Figs. 2-7.)
	
	%TC:ignore
	%TC:endignore
	
	\subsection{Regression model of MPS95 and model interpretation} %correct subscriptions
	Ordinary Least Squares (OLS) was used to build linear regression models to quantify the contributions of kinematic factors (predictors) to the prediction of MPS95 (outcome). The OLS model captures the most direct and explicit predictive power of a factor in terms of explained variance, and therefore it was used to investigate the predictive power of factors for MPS95.
	
	To analyze the relative predictive power of kinematic factors (different derivative order/component and magnitude/power) in linear regression models, four statistical interpretation methods were used: zero-order correlation coefficients \cite{Nathans2012}, structure coefficients \cite{Courville2001}, commonality analysis \cite{CA} and dominance analysis \cite{Budescu1993}. Considering the possible collinearity among features selected, the methods were chosen because they are interpretation methods without the requirement of independence of the predictors. Besides the predictive power analysis of the direct effect of each factor (zero-order correlation coefficients and structure coefficients), the common effects of factors and dominance relationships of factors were also studied by commonality analysis \cite{CA} and dominance analysis \cite{Budescu1993}, respectively.
	
	Zero-order correlation coefficients measure the direct effect by the Pearson correlation between each predictor and the outcome, without considering any other predictors in the OLS regression model \cite{UMN,Nathans2012}. In this study, as the features were grouped to factors, the coefficient of determination ($R^2$) was calculated when each factor was in the regression, in order to quantify the contribution of each factor.
	
	Structure coefficients quantify the direct effect of a predictor by the Pearson correlation between a predictor and the predicted MPS95 value given by the full regression model with all predictors \cite{UMN,Courville2001}. In this study, after we have grouped features into factors, the coefficient of determination ($R^2$) between a kinematic factor and the predicted MPS95 value was calculated to quantify the contribution of each kinematic factor in the regression. 
	
	Commonality analysis (CA) \cite{UMN,CA} is a comprehensive method used to analyze the relative predictive power of kinematic factors in OLS regression, as it takes into consideration not only the direct effect of each predictor, but also the unique ($U$) and common information ($C$) when multiple predictors are present \cite{CA}. Commonality analysis consists of two parts: all-possible-subsets regression and the decomposition of explained variance \cite{Prunier}. To interpret the CA result, the following regression sample can be considered:
	\begin{align}
		Y = \beta_0 + \beta_m m + \beta_n n + \epsilon
	\end{align}
	where the outcome variable $Y$ is modeled by two predictors $m$ and $n$ \cite{CA}. $\beta_m$, $\beta_n$ and $\beta_0$ are parameters of the model and $\epsilon$ is regression model error. The unique information $U$ of predictor $m$ and $n$ are calculated by:
	\begin{align}
		U_m &= R^2_{m, n} - R^2_{n} \\
		U_n &= R^2_{m, n} - R^2_{m}
	\end{align}
	
	where $R^2_{m, n}, R^2_{n}, R^2_{m}$ are the coefficient of determination ($R^2$) of the model $Y = \beta_0 + \beta_m m + \beta_n n + \epsilon$ and the subset models $Y = \beta_n n + \epsilon$ and $Y = \beta_m m + \epsilon$ respectively. The common information $C$ of predictor $m$ and $n$ is calculated by:
	\begin{align}
		C_{m, n} = R^2_{n} + R^2_{m} - R^2_{m,n}
	\end{align}
	The unique information of a factor can be interpreted as the $R^2$ increments by adding the factor into a multiple regression model. The total information of a factor is the $R^2$ of the corresponding regression subset model, which can be explained by the sum of the unique information and common information: for example, the total information of $m$ ($T_m$) is calculated by:
	\begin{align}
		T_m = R^2_m = U_m + C_{m, n}
	\end{align}
	The unique information and the common information are regarded as the commonality coefficients, which denote the part of explained variance ($R^2$) either contributed uniquely by one factor or is shared by more than one factors. Generally, CA decomposes the effect of each factor in a multiple regression into different information with $R^2$ of each subset model.
	
	Dominance analysis (DA) \cite{Budescu1993} provides the analysis of dominance relationships among factors by calculating the increments of $R^2$ when a factor is added to every possible subset models. In DA results, level is a unique term which denotes the number of predictors of a subset model \cite{UMN,Budescu1993}. To understand the result of DA (e.g. Table \ref{tab:da_HM}), level is the number of factors in the first column denoted by $k$. Each entry denotes the incremental $R^2$ contribution of adding a factor to a model at a specific level. Dominance of a factor in level $i$ is determined by averaging all increments in $R^2$ in the level $i$. The factor with the largest averaged contribution dominates over other factors in this level. For example, in the DA of derivative order, the averaged contribution of the first-order features in level 1 ($\Bar{\Delta_{l1} R^2_{o1}}$) is calculated by:
	\begin{align}
		\Bar{\Delta_{l1} R^2_{o1}} = \frac{1}{\binom{3}{1} - 1}[(R^2_{o0, o1} - R^2_{o0}) + (R^2_{o2, o1} - R^2_{o2})]
	\end{align}
	where $\binom{3}{1} - 1$ is the number of subset models evaluated; $R^2_{o0, o1}$ is the $R^2$ of the model that contains zero-order features ($o0$) and first-order features ($o1$); $R^2_{o0}$ is the $R^2$ of the model that contains zero-order features ($o0$). $R^2_{o2, o1}$ and $R^2_{o2}$ can be interpreted in a similar manner.
	The averaged contribution of level 1 is recorded in result tables where the first column is: $k = 1$ averaged. To avoid the computational cost in large numbers of subset models, in the DA of powers, the powers of $1, 0.5, 2, 3$ were selected because the higher powers showed lower predictive power according to zero-order correlation coefficients and structure coefficients.
	
	\subsection{Statistical Tests}
	
	100 iterations of bootstrapping resampling \cite{bootstrapping} were done to test statistical significance. To compare the different factors with zero-order correlation coefficients and structure coefficients, the mean zero-correlation coefficients/structure coefficients of each factor in the bootstrapping were normalized to the sum of mean zero-correlation coefficients/structure coefficients of all factors. According to the bootstrapping results, paired t-tests between the 100 zero-correlation coefficients/structure coefficients given by most predictive factor and the 100 zero-correlation coefficients/structure coefficients given by the second most predictive factor found on each dataset were done to test the statistical significance. Because commonality analysis and dominance analysis require the number of samples to be larger than the number of features to ensure validity and interpretability, the datasets NFL and NHTSA were excluded in these two analyses.
	
	\section{Results} 
	\subsection{Predictive power of different derivative orders}
\begin{figure}
		\centering
		\includegraphics[
		page=1,
		width=\linewidth,
		keepaspectratio,
		trim=0 0 0 0pt
		]{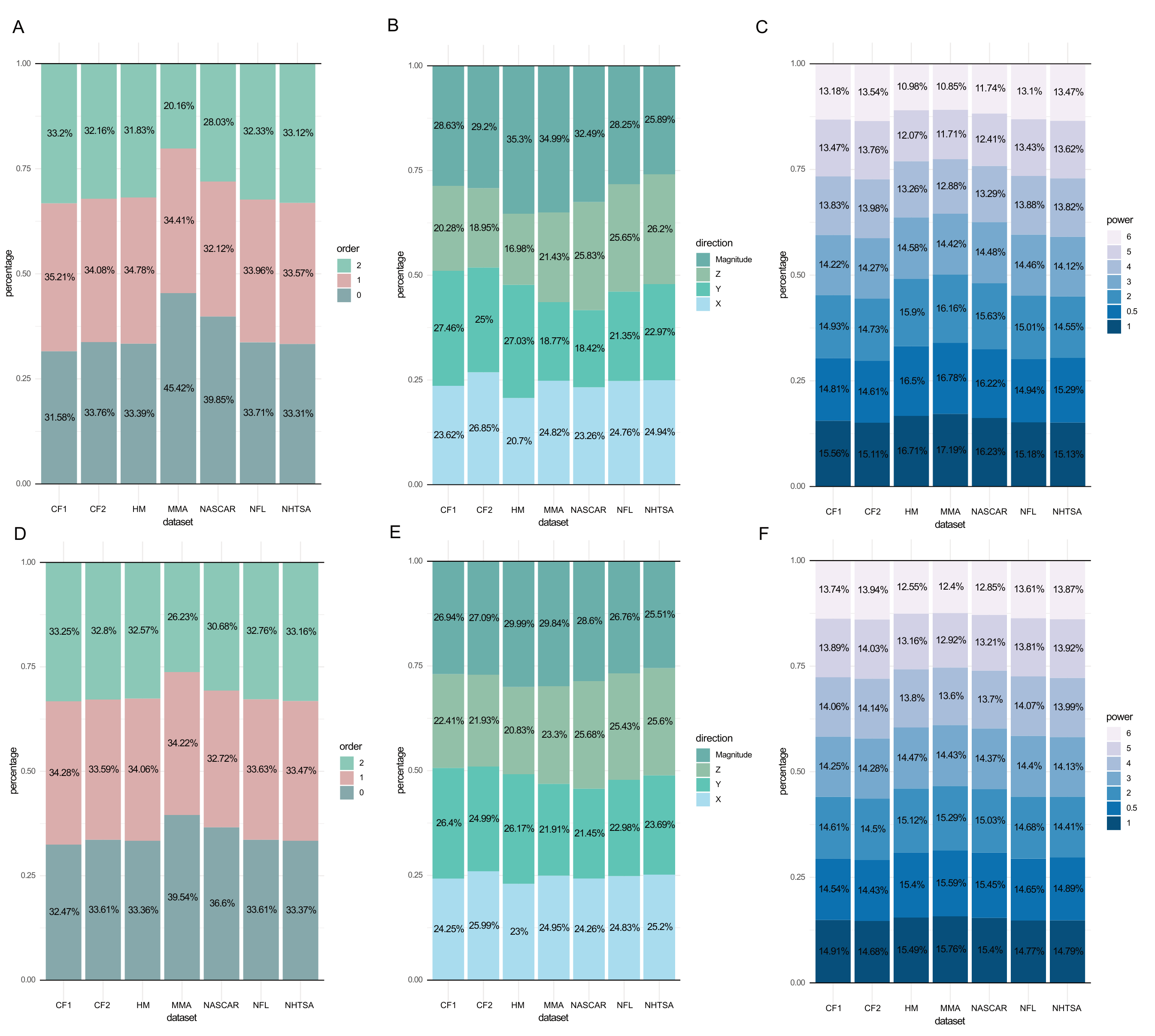}
		\caption[short caption]{Stacked bar plots of the relative variance explained by different factors based on normalized mean zero-correlation coefficients/structure coefficients of kinematic factors in the regression of MPS95. The mean zero-correlation coefficients/structure coefficients were firstly calculated in 100 iterations of bootstrapping resampling and the results were normalized by the sum of the mean zero-correlation coefficients/structure coefficients of different factors. A-C. Percentage contributions of features of three different derivative orders (0, 1, 2), features of four different kinematic components (x-axis, y-axis, z-axis, magnitude), features of seven different powers (1, 0.5, 2, 3, 4, 5, 6) given by zero-order correlation coefficients. D-F. Percentage contributions of features of three different derivative orders, features of four different kinematic components, features of seven different powers given by structure coefficients.}
		\label{fig:mps_percentage.pdf}
	\end{figure}

	The predictive power of three factors of derivative orders were analyzed using the methods mentioned in Section 2.3 (Fig. \ref{fig:mps_percentage.pdf} A and D). According to both zero-order correlation coefficients and structure coefficients, on all datasets but dataset MMA and NASCAR, the first-order features (angular acceleration) were the most predictive features while the second-order features (angular jerk) were the least predictive features ($p<0.05$, see details in section 2.4), with an exception on dataset CF2 that the structure coefficients found no statistical significant difference between the angular acceleration and the angular velocity ($p=0.60$). On dataset MMA and NASCAR, both statistical methods showed zero-order features (angular velocity) were the most predictive ($p<0.05$). What must be mentioned here is that the numbers in Fig. \ref{fig:mps_percentage.pdf} can be interpreted as the relative explained variance because we have normalized the results given by zero-order correlation coefficients and structure coefficients after bootstrapping each dataset for 100 times. Furthermore, it can be shown by the results that even the least predictive factor was also predictive of MPS95.
	
\begin{sidewaysfigure}
		\centering
		\includegraphics[
		page=1,
		width=\linewidth,
		keepaspectratio,
		trim=0 0 0 0pt
		]{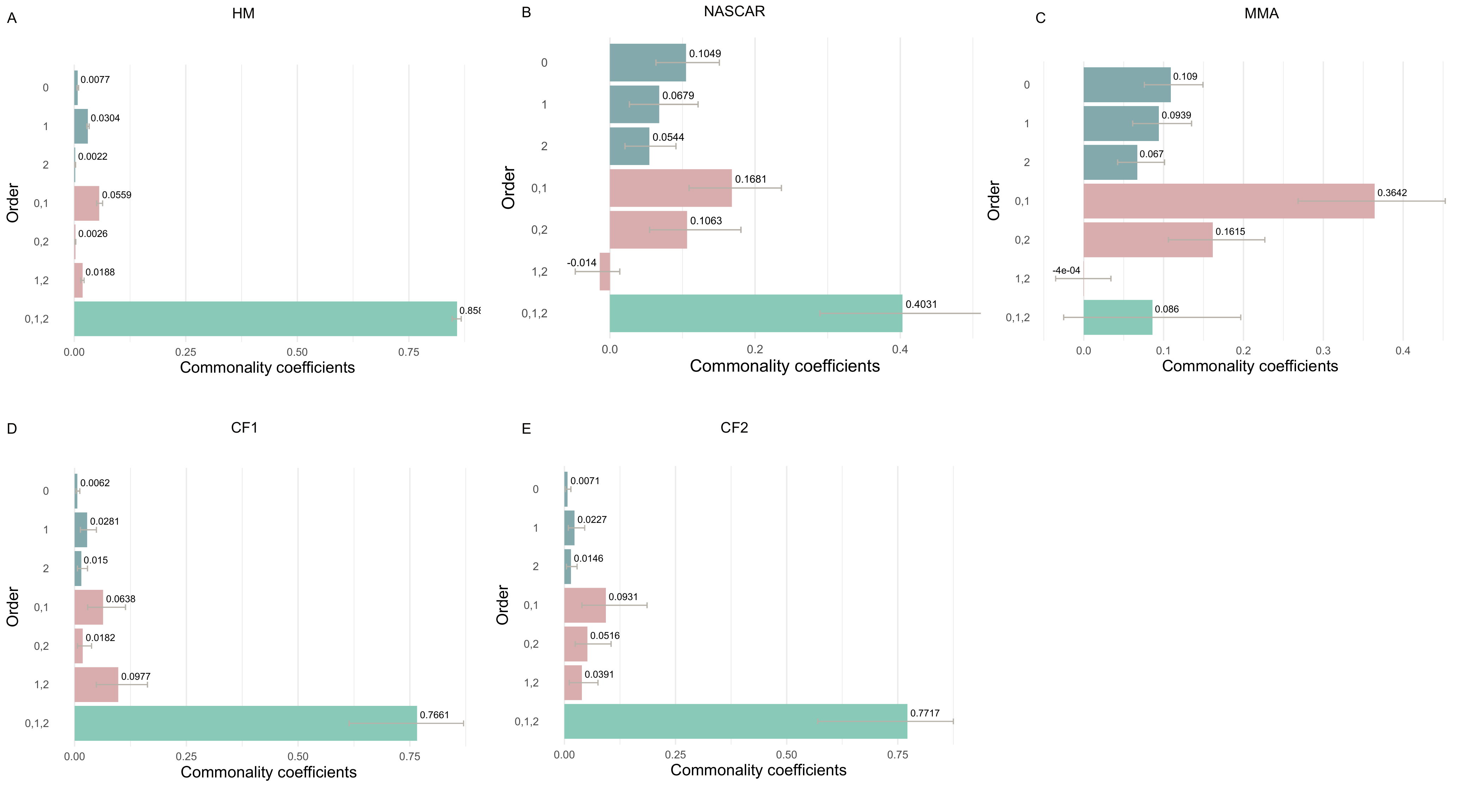}
		\caption[short caption]{Commonality analysis results for unique or common information of different predictors in the regression of MPS95 from features of three different derivative orders (0,1,2) on five different datasets. Positive commonality coefficients indicate that a part of the explained variance ($R^2$) of the dependent variable is either unique to certain factor or shared between some factors. Negative commonality coefficients indicate that there is a suppressor effects between some factors.}
		\label{fig:ca_mps_order}
	\end{sidewaysfigure}

	Commonality analysis showed that the common information of all three factors had the highest predictive power on datasets HM, CF1, CF2 and NASCAR (Fig. \ref{fig:ca_mps_order}). On dataset MMA, the common information of zero-order features and first-order features had the highest predictive power. The dominance analysis showed that on dataset HM, CF1, CF2, the first-order features dominated in every level (Table \ref{tab:da_HM}, Supplementary Tables 1-4) and on dataset MMA and NASCAR, the zero-order features dominated in every level.
	
	\subsection{Predictive power of different kinematic components}
	The same interpretation methods were used to analyze the four kinematic components: the three directions and the magnitude (Fig. \ref{fig:mps_percentage.pdf} B and E). According to both zero-order correlation coefficients and structure coefficients, on all datasets but dataset NHTSA, magnitude features were the most predictive features ($p<0.05$, see details in section 2.4). Zero-order correlation coefficients showed the features of z-axis component were the most predictive on dataset NHTSA ($p<0.05$). This finding was not statistically significant according to structure coefficients ($p=0.17$). Among the components in the three directions, both interpretation methods showed that the ranking of predictive power was not the same across datasets. 
	
\begin{sidewaysfigure}
		\centering
		\includegraphics[
		page=1,
		width=\linewidth,
		keepaspectratio,
		trim=0 0 0 0pt
		]{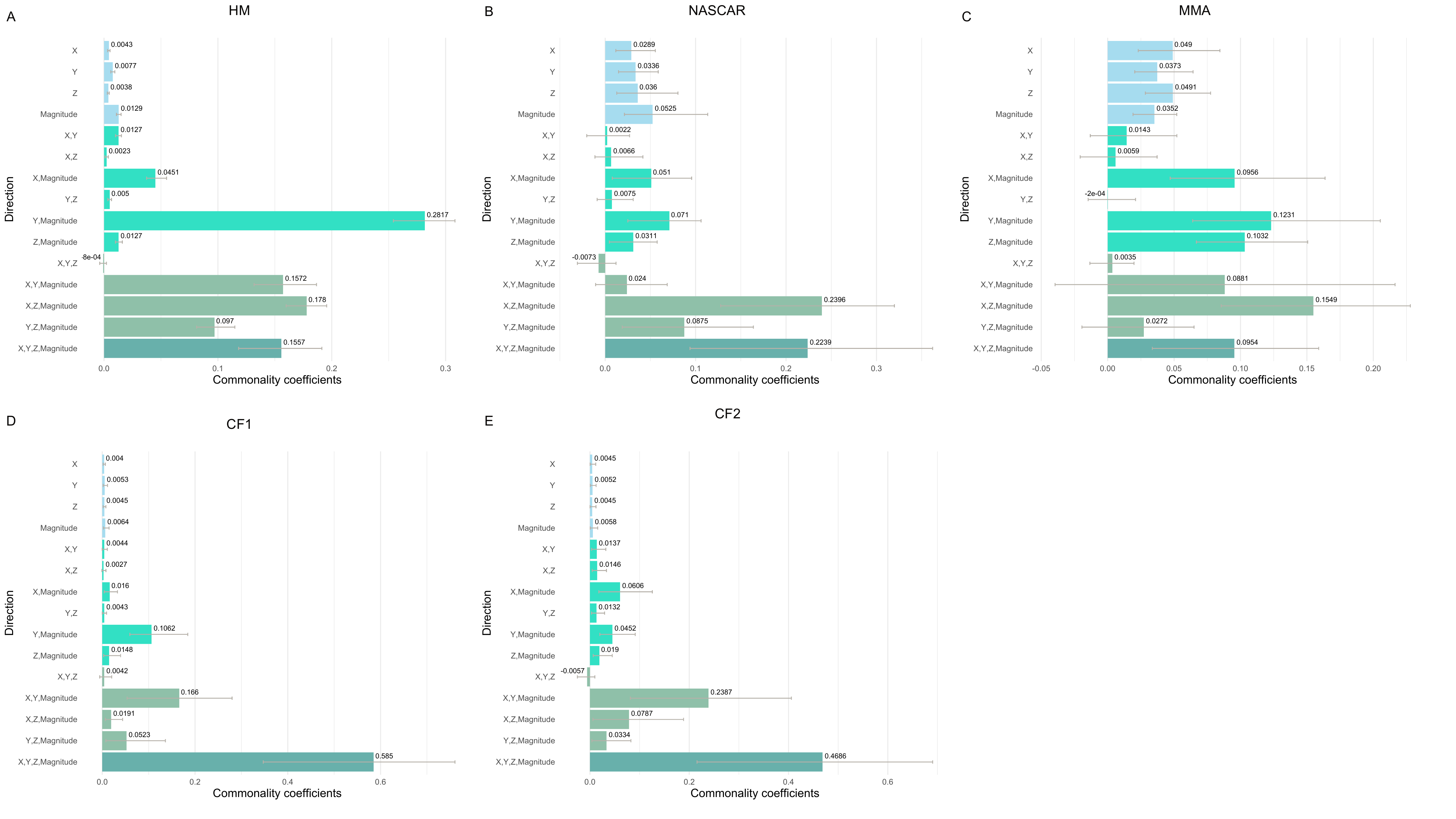}
		\caption[short caption]{Commonality analysis results for unique or common information of different predictors in the regression of MPS95 from features of four different kinematic components (x-axis, y-axis, z-axis and magnitude) on five different datasets. Positive commonality coefficients indicate that a part of the explained variance ($R^2$) of the dependent variable is either unique to certain factor or shared between some factors. Negative commonality coefficients indicate that there is a suppressor effects between some factors.}
		\label{fig:ca_mps_dir}
	\end{sidewaysfigure}

	Commonality analysis showed the predictive power varied with different datasets (Fig. \ref{fig:ca_mps_dir}) but the most predictive information always involved magnitude. On dataset HM, the common information of features from y-axis and the magnitude showed the highest predictive power. On dataset CF1 and CF2, the common information of features from x-axis, y-axis, z-axis and the magnitude was the most predictive. On dataset MMA, the common information of features from x-axis, z-axis, and the magnitude was the most predictive. Dominance analysis showed that magnitude features dominated in the majority of levels of analyses with one exception on dataset MMA and NASCAR, where x-axis features dominated over other features on level 3 (where the features from y-axis, z-axis, and the magnitude were already included)(Table \ref{tab:da_HM}, Supplementary Tables 1-4).
	
	\begin{table}[htbp]
		\centering
		\small
		\caption{Dominant analysis results for dataset HM.}
		\hspace*{-2cm}
		\begin{subtable}[t]{.49\textwidth}
			\centering
			\tiny
			\caption{Dominance analysis of different derivative orders on dataset HM.}
			\begin{tabular}{p{6.5em}cccc}
				\toprule
				\multirow{2}[4]{*}{\textbf{Order}} & \multicolumn{1}{c}{\multirow{2}[4]{*}{\textbf{Mean R2}}} & \multicolumn{3}{c}{\textbf{Additional Contribution}} \\
				\cmidrule{3-5}    \multicolumn{1}{c}{} &   & \textbf{0} & \textbf{1} & \textbf{2} \\
				\midrule
				\textbf{k=0 average} & 0.0000 & 0.9240 & 0.9630 & 0.8811 \\
				\multicolumn{1}{l}{0} & 0.9240 & 0.0000 & 0.0493 & 0.0213 \\
				\multicolumn{1}{l}{1} & 0.9630 & 0.0103 & 0.0000 & 0.0048 \\
				\multicolumn{1}{l}{2} & 0.8811 & 0.0643 & 0.0867 & 0.0000 \\
				\textbf{k=1 average} & 0.0000 & 0.0373 & 0.0680 & 0.0131 \\
				0,1 & 0.9733 & 0.0000 & 0.0000 & 0.0022 \\
				0,2 & 0.9454 & 0.0000 & 0.0301 & 0.0000 \\
				1,2 & 0.9678 & 0.0077 & 0.0000 & 0.0000 \\
				\textbf{k=2 average} & 0.0000 & 0.0077 & 0.0301 & 0.0022 \\
				0,1,2 & 0.9755 & 0.0000 & 0.0000 & 0.0000 \\
				\bottomrule
			\end{tabular}
		\end{subtable}
		\hspace*{1cm}
		\begin{subtable}[t]{.48\textwidth}
			\centering
			\tiny
			\caption{Dominance analysis of different powers on dataset HM.}
			\begin{tabular}{lccccr}
				\toprule
				\multicolumn{1}{c}{\multirow{2}[2]{*}{\textbf{Power}}} & \multirow{2}[2]{*}{\textbf{Mean R2}} & \multicolumn{4}{c}{\textbf{Additional Contributions}} \\
				&   & \textbf{1} & \textbf{0.5} & \textbf{2} & \textbf{3} \\
				\midrule
				\textbf{k=0 average} & 0 & 0.9664 & 0.9545 & 0.9201 & 0.8437 \\
				1 & 0.9664 & 0 & 0.0023 & 0.0028 & 0.0029 \\
				0.5 & 0.9545 & 0.0142 & 0 & 0.0126 & 0.0108 \\
				2 & 0.9201 & 0.049 & 0.0469 & 0 & 0.0482 \\
				3 & 0.8437 & 0.1256 & 0.1216 & 0.1247 & 0 \\
				\textbf{k=1 average} & 0 & 0.0629 & 0.0569 & 0.0467 & 0.0206 \\
				1, 0.5 & 0.9686 & 0 & 0 & 0.0021 & 0.0024 \\
				1, 2 & 0.9691 & 0 & 0.0016 & 0 & 0.0023 \\
				1, 3 & 0.9693 & 0 & 0.0018 & 0.0021 & 0 \\
				0.5, 2 & 0.9671 & 0.0037 & 0 & 0 & 0.004 \\
				0.5, 3 & 0.9652 & 0.0058 & 0 & 0.0058 & 0 \\
				2, 3 & 0.9684 & 0.0031 & 0.0027 & 0 & 0 \\
				\textbf{k=2 average} & 0 & 0.0042 & 0.002 & 0.0034 & 0.0029 \\
				1, 0.5, 2 & 0.9708 & 0 & 0 & 0 & 0.0016 \\
				1, 0.5, 3 & 0.971 & 0 & 0 & 0.0013 & 0 \\
				1, 2, 3 & 0.9714 & 0 & 0.0009 & 0 & 0 \\
				0.5, 2, 3 & 0.971 & 0.0013 & 0 & 0 & 0 \\
				\textbf{k=3 average} & 0 & 0.0013 & 0.0009 & 0.0013 & 0.0016 \\
				1, 0.5, 2, 3 & 0.9723 & 0 & 0 & 0 & 0 \\
				\bottomrule
			\end{tabular}%
		\end{subtable}
		\begin{subtable}{1\textwidth}
			\centering
			\tiny
			\caption{Dominance analysis of different components on dataset HM.}
			\hspace*{-3cm}
			\begin{tabular}{lccccc}
				\toprule
				\multicolumn{1}{c}{\multirow{2}[2]{*}{\textbf{Directions}}} & \multirow{2}[2]{*}{\textbf{Mean R2}} & \multicolumn{4}{c}{\textbf{Additional Contributions}} \\
				&   & \textbf{X} & \textbf{Y} & \textbf{Z} & \textbf{Magnitude} \\
				\midrule
				\textbf{k=0 average} & 0 & 0.5544 & 0.7182 & 0.4525 & 0.9406 \\
				X & 0.5544 & 0 & 0.3917 & 0.1196 & 0.4047 \\
				Y & 0.7182 & 0.2279 & 0 & 0.1948 & 0.2467 \\
				Z & 0.4525 & 0.2216 & 0.4605 & 0 & 0.4984 \\
				Magnitude & 0.9406 & 0.0185 & 0.0243 & 0.0103 & 0 \\
				\textbf{k=1 average} & 0 & 0.156 & 0.2922 & 0.1082 & 0.3833 \\
				X,Y & 0.9461 & 0 & 0 & 0.0162 & 0.0255 \\
				X,Z & 0.674 & 0 & 0.2883 & 0 & 0.2938 \\
				X,Magnitude & 0.9591 & 0 & 0.0125 & 0.0087 & 0 \\
				Y,Z & 0.913 & 0.0494 & 0 & 0 & 0.0581 \\
				Y,Magnitude & 0.9649 & 0.0067 & 0 & 0.0062 & 0 \\
				Z,Magnitude & 0.9509 & 0.017 & 0.0202 & 0 & 0 \\
				\textbf{k=2 average} & 0 & 0.0243 & 0.107 & 0.0104 & 0.1258 \\
				X,Y,Z & 0.9623 & 0 & 0 & 0 & 0.013 \\
				X,Y,Magnitude & 0.9716 & 0 & 0 & 0.0037 & 0 \\
				X,Z,Magnitude & 0.9679 & 0 & 0.0075 & 0 & 0 \\
				Y,Z,Magnitude & 0.9711 & 0.0042 & 0 & 0 & 0 \\
				\textbf{k=3 average} & 0 & 0.0042 & 0.0075 & 0.0037 & 0.013 \\
				X,Y,Z,Magnitude & 0.9754 & 0 & 0 & 0 & 0 \\
				\bottomrule
			\end{tabular}\hspace*{-3cm}%
		\end{subtable}
		\label{tab:da_HM}
	\end{table}%
	
	\subsection{Predictive power of different powers}
	In the power analysis (Fig. \ref{fig:mps_percentage.pdf} C and F), the relative relationship of the variance explained by 7 different powers were shown. Both the zero-order correlation coefficients and the structure coefficients showed that, on all datasets except NHTSA, the first-power features were the most predictive features ($p<0.05$, see details in section 2.4) with an exception on dataset NASCAR, where both methods found no statistical significance between the first-power features and square-root features ($p>0.05$). However, the square-root features were the most predictive feature on the dataset NHTSA ($p<0.05$). The powers smaller than 3 were significantly more predictive than the remaining powers. Besides the power of 0.5, 1, and 2, as the power increases, the features showed gradually decreasing predictive power in the regression, although the differences were small.
	
\begin{sidewaysfigure}
		\centering
		\includegraphics[
		page=1,
		width=\linewidth,
		keepaspectratio,
		trim=0 0 0 0pt
		]{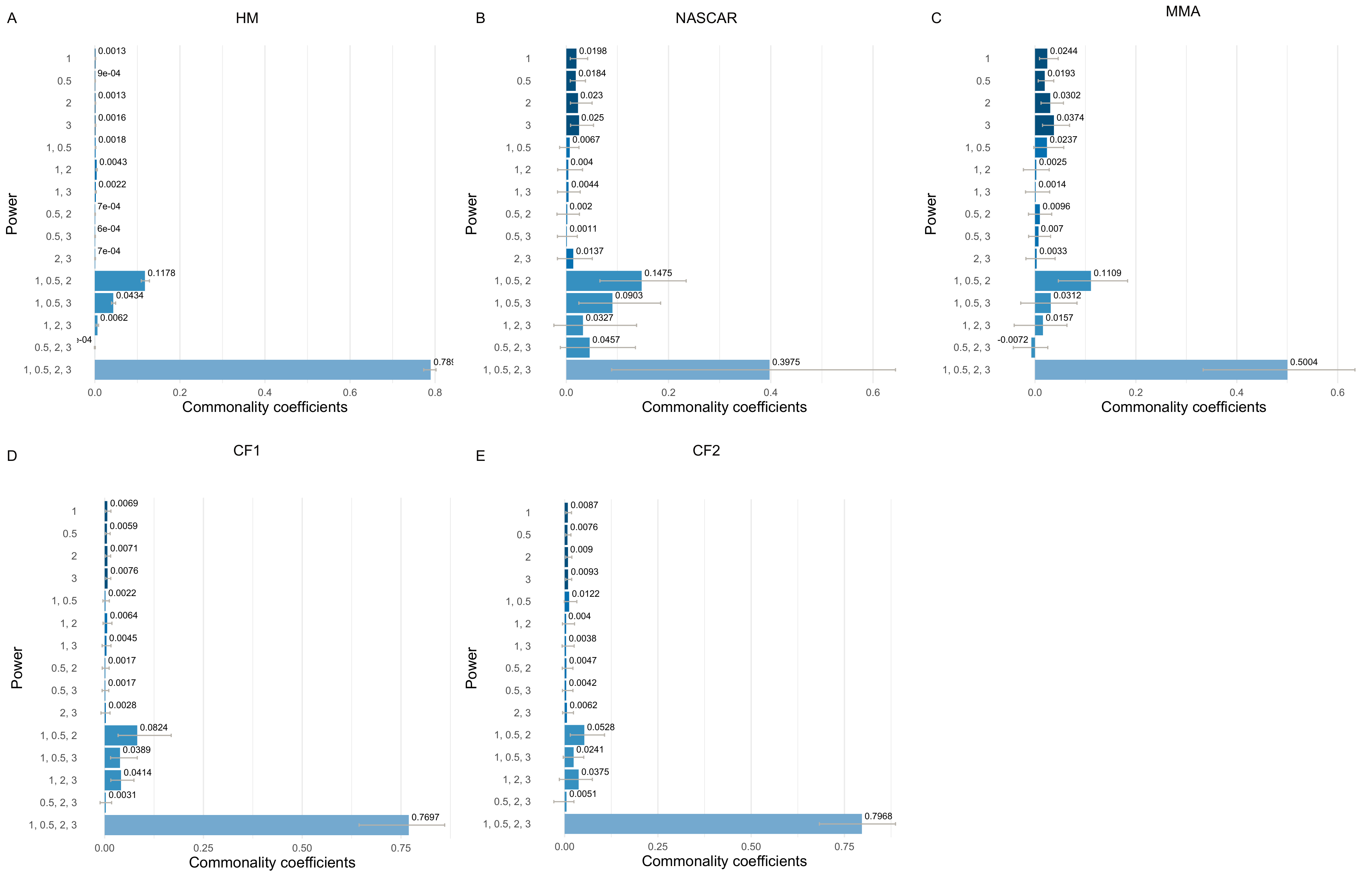}
		\caption[short caption]{Commonality analysis results for unique or common information of different predictors in the regression of MPS95 from features of four different powers (1, 0.5, 2, 3) on five different datasets. Positive commonality coefficients indicate that a part of the explained variance ($R^2$) of the dependent variable is either unique to certain factor or shared between some factors. Negative commonality coefficients indicate that there is a suppressor effects between some factors.}
		\label{fig:ca_mps_power}
	\end{sidewaysfigure}

	In commonality analysis (Fig. \ref{fig:ca_mps_power}) and dominance analysis (Table \ref{tab:da_HM}, Supplementary Tables 1-4), the factors were limited to be of power 1, 0.5, 2, and 3, because the factor contribution gradually decreased with increasing power and a larger number of factor subsets rapidly increased the number of result entries, which made it difficult to interpret. According to the results of commonality analysis, the common information of all factors had the highest predictive power, which indicates that changing the power of features does not provide new information in the regression. The dominance analysis showed that the first-power features dominates over each level, with exceptions on dataset MMA and dataset NASCAR where different factors showed dominance on different levels, which indicates that the predictive power of different factors were not clearly ranked on dataset MMA and dataset NASCAR. 
	
	\section{Discussion}
	Although many BIC \cite{HIC,BrIC}, physics-based models \cite{BAM,Gabler2019} and deep-learning-based models \cite{Shaoju,MLHM,Ji20} have been developed to accurately estimate MPS95, they rely on black-box models or strong mechanical and mathematical model assumptions. There is a lack of systematic analyses over the predictive power of different kinematics in estimating MPS95 in a data-driven manner. This study applied four statistical interpretation methods to analyze the regression of MPS95 to explicitly evaluate the predictive power of different factors of rotational parameters of head impact kinematics. In terms of the predictive power among factors when viewed individually (the direct effect defined by Budescu \cite{Budescu1993}), the zero-order correlation coefficients and structure coefficients manifest that the following kinematic factors generally show the highest predictive power: 1) features based on the first derivative order of angular velocity (angular acceleration), 2) features based on the magnitude of the kinematics and 3) features of power 1 (Fig. \ref{fig:mps_percentage.pdf}). These findings were generally supported by the dominance analysis in which these three factors dominated in most levels of analyses (Table \ref{tab:da_HM}, Supplementary Tables 1-4). The finding that magnitude features were generally more predictive of MPS95 than any individual spatial directions supports the majority of BIC designs, which take the magnitude of kinematics into consideration \cite{HIC,RIC,CP}. While the angular velocity and angular jerk features are predictive, the angular acceleration features generally showed higher predictive power on most datasets, which agrees with the findings in \cite{InertialF}. Several BIC incorporate the magnitude of the angular acceleration into their respective computational models, such as RIC \cite{RIC} and Power Rotation Head Injury Criterion (PRHIC) \cite{PRHIC}. The first-power features were the most predictive features on most datasets. This generally shows that the relationship between the predictor (kinematic features) and the outcome (MPS95) may tend to approximate a linear relationship rather than a higher order polynomial relationship. 
	
	There were several dissimilarities across the datasets. First, both zero-order correlation coefficients and structure coefficients showed that angular velocity features exhibit statistically significantly higher predictive power on dataset MMA and dataset NASCAR (Fig. \ref{fig:mps_percentage.pdf}). The dominance analysis also supported the findings that the zero-order features dominated in all levels on dataset MMA and dataset NASCAR while the first-order features (angular acceleration) dominated on the other datasets. The higher predictive power of the angular velocity on MMA impacts may be related to the fact that the duration of MMA impacts is generally shorter than their counterparts in other datasets. For short-duration impacts, angular velocity tends to be more predictive of brain strain \cite{Panzer1,InertialF}. It should be noted that the change of angular velocity was considered to be more correlated to the brain strain \cite{RIC}, and the zero-order features are the estimation of the change of angular velocity. Based on this finding, BrIC \cite{BrIC}, kinematic rotational brain injury criterion (BRIC) \cite{BRIC}, and rotational velocity change index (RVCI) \cite{RVCI}, which all take into consideration the peak values of angular velocity or the numerical integral of angular acceleration, may be more effective in predicting MPS95 on the short-duration head impacts, such as those experienced in MMA. This finding agrees with the evaluation of BIC by Zhan et al. \cite{BIC} in a recent study that found BrIC, BRIC, RVCI, and the maximum value of angular velocity magnitude were among the top five (out of 18 total) most accurate BIC in predicting MPS95 on MMA head impacts. On the two football datasets (CF1 and CF2), besides the most predictive magnitude features, the x-axis features and y-axis features show higher predictive power than the z-axis features. This finding suggests a potential explanation that the side or rear impacts are more likely to lead to obvious performance decrement (OPD) of the players \cite{Bartsch2020}. 
	
	Furthermore, on dataset NHTSA, the square-root features showed higher predictive power than the first-power features, which were the most predictive on the other datasets. On dataset NASCAR, there was no statistical significance between square-root features and first-power features. This finding indicates that the underlying relationship between the features and MPS95 may be slightly different among various datasets (i.e., different types of head impacts). This may also raise concerns of interchangeably using BIC designed for different types of head impacts, which is investigated by Zhan et al. \cite{BIC}. Extra model validation may be needed to show the transferability of certain BIC when applied to different types of head impacts.
	
	Additionally, commonality analysis provided more information of the predictive power unique and common to different factors. In the derivative order analysis, the zero-order features were the most predictive factor on dataset MMA by both zero-order correlation coefficients and structure coefficients. Furthermore, commonality analysis showed that on dataset MMA, the common information of zero-order and first-order features was the most predictive, which indicates that the common information of these two factors explained a lot of variances. These two findings collectively indicate that the subtle, unique information of the zero-order features leads the zero-order features to be the most predictive, while the majority of predictive power is provided by the common information that are mutual to both factors. In the kinematic component analysis (Fig. \ref{fig:ca_mps_dir}), commonality analysis showed that the most predictive information was in the common information provided by y-axis features and the magnitude features for dataset HM, by x-axis, y-axis, z-axis and the magnitude features for dataset CF1 and CF2, and by x-axis, z-axis and the magnitude features for dataset MMA. These findings indicate the predictive information of MPS95 may come from different spatial directions in different types of head impacts. To accurately predict MPS95, not only the magnitude but also the other spatial directions are predictive, while the predictive power of the kinematics component varies with types of sports. This finding indicates that the designs of BIC which take different spatial directions into consideration, such as in BAM \cite{BAM}, BrIC \cite{BrIC} and RVCI \cite{RVCI}, may bear an advantage on some types of head impacts and provide more accurate brain injury risk evaluation.
	
	It should be noted in this study that we did not directly interpret the values or the absolute values of the regression coefficients in the regression model because there can be strong collinearity among the features we selected. The correlated predictors in the regression model may not only lead to ambiguous interpretation of linearly correlated predictors but also cause a variation inflation effect \cite{VIF} that leads to unstable regression coefficients. The inflated variance of regression coefficients can be hard to interpret as we took the mean regression coefficient after bootstrapping the datasets in 100 iterations to ensure the robustness of the analysis. 
	
	Based on the findings in this study, new mathematical forms can be designed and validated to evaluate brain injury risk in a more accurate manner. For instance, brain injury criteria can take into consideration both angular acceleration and angular velocity \cite{Panzer1,BRIC}, the magnitude as well as the components in different spatial directions \cite{BrIC,BAM,RVCI}, and the first-power as well as the square-root math forms for better risk evaluation in a wider range of head impacts.
	
	Although this study finds the difference in the predictive power among factors of head impact kinematics, there are limitations to these conclusions. Firstly, the features we analyzed are limited in the power and the derivative order. That is because we deemed that the higher derivative orders were highly likely influenced by lower signal-to-noise ratio caused by numerical difference. According to the present study, as the power increases when the order is higher than 2, the predictive power generally decreases. It may be possible that the higher derivative orders and powers show unexpected higher predictive power in MPS95 regression. 
	
	Secondly, we applied the ordinary linear regression model to investigate the most direct linear relationship between predictors and the outcome. Although several polynomial features were included to account for a certain degree of non-linearity, we did not consider other non-linearity forms such as sinusoidal, logarithmic, exponential, etc., because any non-linear function can be approximated by polynomials. It is possible that other types of non-linearity can better characterize the relationship between kinematics and brain strain. 
	
	Additionally, as the KTH model was used as the finite element head model (FEHM), it is simplified when compared to the state-of-the-art FEHMs \cite{Zhao19,Li20,Fahlsted21}. Without a separation in the normal direction between subdural cerebral spinal fluid and brain, it has tangential sliding. No gyri or sulci which have been confirmed to have a significant influence on the FEHM behavior and its intracranial motion, are not included in the KTH model. In the future, the more recently developed FEHMs can be applied to validate the results.
	
	Furthermore, this study is a data-driven statistical analysis and the findings of this study may not be biomechanically important deemed by experts. Finally, while various measures of local brain deformation are used to assess the brain injury risk, no consensus has been reached yet about the best metric to describe the tissue strains and discriminate the injury severity \cite{Zhou21}. The current study focused on the relationship between the kinematic predictor and MPS, this can be extended to other strain-based metrics, such as tract-oriented strain, CSDM, etc.
	
	This work applied four different statistical interpretation methods to analyze the linear regression of MPS95 on features extracted from the kinematics in different types of head impacts. The major findings of this work show that the features based on 1) angular acceleration, 2) the magnitude, and 3) the first power are generally shown to be the most predictive features on the majority of head impact datasets. There are also dissimilarities across datasets, such as features based on angular velocity are the most predictive features on dataset MMA and NASCAR and features based on square-root power are the most predictive features on dataset NHTSA. The predictive information from different spatial directions of kinematics varies with different types of head impacts. The analysis presented in this study can help the development of brain injury criteria.
	
	\section{Acknowledgements}
	This research was supported by the Pac-12 Conference’s Student-Athlete Health and Well-Being Initiative, the National Institutes of Health (R24NS098518), Taube Stanford Children’s Concussion Initiative and Stanford Department of Bioengineering. The authors also want to express their gratitude for the insights from the linear model course (STATS305A) offered by Dr. Trevor Hastie at Stanford University.
	
	\clearpage
	\newpage
	\bibliographystyle{unsrt}

\begin{thebibliography}{9}
		\bibitem{Spencer16} James, Spencer L., et al. "Global, regional, and national burden of traumatic brain injury and spinal cord injury, 1990–2016: a systematic analysis for the Global Burden of Disease Study 2016." The Lancet Neurology 18.1 (2019): 56-87.
		\bibitem{Montenigro17} Montenigro, Philip H., et al. "Cumulative head impact exposure predicts later-life depression, apathy, executive dysfunction, and cognitive impairment in former high school and college football players." Journal of neurotrauma 34.2 (2017): 328-340.
		\bibitem{OKeeffe19} O'Keeffe, Eoin, et al. "Dynamic Blood–Brain Barrier Regulation in Mild Traumatic Brain Injury." Journal of neurotrauma 37.2 (2020): 347-356.
		\bibitem{guiza2017} Güiza, Fabian, et al. "Early detection of increased intracranial pressure episodes in traumatic brain injury: External validation in an adult and in a pediatric cohort." Critical care medicine 45.3 (2017): e316-e320.
		\bibitem{Panzer1} Gabler, Lee F., et al. "Development of a single-degree-of-freedom mechanical model for predicting strain-based brain injury responses." Journal of biomechanical engineering 140.3 (2018).
		\bibitem{Gabler2019} Gabler, Lee F., Jeff R. Crandall, and Matthew B. Panzer. "Development of a second-order system for rapid estimation of maximum brain strain." Annals of biomedical engineering 47.9 (2019): 1971-1981.
		\bibitem{BAM} Laksari, Kaveh, et al. "Multi-directional dynamic model for traumatic brain injury detection." Journal of Neurotrauma 37.7 (2020): 982-993.
		\bibitem{Shaoju} Wu, Shaoju, et al. "Convolutional neural network for efficient estimation of regional brain strains." Scientific reports 9.1 (2019): 1-11.
		\bibitem{MLHM} Zhan, Xianghao, et al. "Rapid Estimation of Entire Brain Strain Using Deep Learning Models." IEEE Transactions on Bio-medical Engineering (2021).
		\bibitem{Ji20} Ghazi, Kianoosh, et al. "Instantaneous Whole-brain Strain Estimation in Dynamic Head Impact." Journal of Neurotrauma ja (2020).
		\bibitem{HIC} Versace, John. A review of the severity index. No. 710881. SAE Technical Paper, 1971.
		\bibitem{BrIC} Takhounts, Erik G., et al. Development of brain injury criteria (BrIC). No. 2013-22-0010. SAE Technical Paper, 2013.
		\bibitem{RIC} Kimpara, Hideyuki, and Masami Iwamoto. "Mild traumatic brain injury predictors based on angular accelerations during impacts." Annals of biomedical engineering 40.1 (2012): 114-126.
		\bibitem{Bain00} Bain, Allison C., and David F. Meaney. "Tissue-level thresholds for axonal damage in an experimental model of central nervous system white matter injury." Journal of biomechanical engineering 122.6 (2000): 615-622.
		\bibitem{Bar-Kochba16} Bar-Kochba, Eyal, et al. "Strain and rate-dependent neuronal injury in a 3D in vitro compression model of traumatic brain injury." Scientific Reports 6.1 (2016): 1-11.
		\bibitem{Cater06} Cater, Heather L., Lars E. Sundstrom, and Barclay Morrison III. "Temporal development of hippocampal cell death is dependent on tissue strain but not strain rate." Journal of biomechanics 39.15 (2006): 2810-2818.
		\bibitem{Donat21} Donat, Cornelius K., et al. "From biomechanics to pathology: predicting axonal injury from patterns of strain after traumatic brain injury." Brain 144.1 (2021): 70-91.
		\bibitem{Fahlstedt15} Fahlstedt, Madelen, et al. "Correlation between injury pattern and finite element analysis in biomechanical reconstructions of traumatic brain injuries." Journal of biomechanics 48.7 (2015): 1331-1335.
		\bibitem{Gennarelli89} Gennarelli, T. A., et al. "Axonal injury in the optic nerve: a model simulating diffuse axonal injury in the brain." Journal of neurosurgery 71.2 (1989): 244-253.
		\bibitem{Hajiaghamemar21} Hajiaghamemar, Marzieh, and Susan S. Margulies. "Multi-scale white matter tract embedded brain finite element model predicts the location of traumatic diffuse axonal injury." Journal of Neurotrauma 38.1 (2021): 144-157.
		\bibitem{Hernandez19} Hernandez, Fidel, et al. "Lateral impacts correlate with falx cerebri displacement and corpus callosum trauma in sports-related concussions." Biomechanics and modeling in mechanobiology 18.3 (2019): 631-649.
		\bibitem{Kang15} Kang, Woo Hyeun, and Barclay Morrison. "Functional tolerance to mechanical deformation developed from organotypic hippocampal slice cultures." Biomechanics and modeling in mechanobiology 14.3 (2015): 561-575.
		\bibitem{McAllister12} McAllister, Thomas W., et al. "Maximum principal strain and strain rate associated with concussion diagnosis correlates with changes in corpus callosum white matter indices." Annals of biomedical engineering 40.1 (2012): 127-140.
		\bibitem{InertialF} Liu, Yuzhe, et al. "Theoretical and numerical analysis for angular acceleration being determinant of brain strain in mTBI." arXiv preprint arXiv:2012.13507 (2020).
		\bibitem{BIC} Zhan, Xianghao, et al. "The relationship between brain injury criteria and
		brain strain across different types of head impacts can be different." Journal of Royal Society Interface 18 (2021): 20210260.
		\bibitem{con1} Sarkar, Shamik, Santanu Majumder, and Amit Roychowdhury. "Factors affecting diffuse axonal injury under blunt impact and proposal for a head injury criteria: a finite element analysis." Critical Reviews™ in Biomedical Engineering 46.4 (2018).
		\bibitem{con2} Weaver, Ashley A., Kerry A. Danelson, and Joel D. Stitzel. "Modeling brain injury response for rotational velocities of varying directions and magnitudes." Annals of biomedical engineering 40.9 (2012): 2005-2018.
		\bibitem{con3} Tiernan, Stephen, and Gary Byrne. "The effect of impact location on brain strain." Brain injury 33.4 (2019): 427-434.
		\bibitem{con4} Tse, Kwong Ming, et al. "Effect of helmet liner systems and impact directions on severity of head injuries sustained in ballistic impacts: a finite element (FE) study." Medical \& biological engineering \& computing 55.4 (2017): 641-662.
		\bibitem{con5} Post, Andrew, et al. "Traumatic brain injuries: the influence of the direction of impact." Neurosurgery 76.1 (2015): 81-91.
		\bibitem{Holbourn} Holbourn, A. H. S. "Mechanics of head injuries." The Lancet 242.6267 (1943): 438-441.
		\bibitem{Kleiven2007} Ho, Johnson, and Svein Kleiven. "Dynamic response of the brain with vasculature: a three-dimensional computational study." Journal of biomechanics 40.13 (2007): 3006-3012.
		\bibitem{DummyFEA} Giudice, J. Sebastian, et al. "Development of open-source dummy and impactor models for the assessment of American football helmet finite element models." Annals of biomedical engineering 47.2 (2019): 464-474.
		\bibitem{camarillo2013} Camarillo, David B., et al. "An instrumented mouthguard for measuring linear and angular head impact kinematics in American football." Annals of biomedical engineering 41.9 (2013): 1939-1949.
		\bibitem{MGValidation}Liu, Yuzhe, et al. "Validation and Comparison of Instrumented Mouthguards for Measuring Head Kinematics and Assessing Brain Deformation in Football Impacts." arXiv preprint arXiv:2008.01903 (2020).
		\bibitem{MMA2} Tiernan, Stephen, et al. "Concussion and the severity of head impacts in mixed martial arts." Proceedings of the Institution of Mechanical Engineers, Part H: Journal of engineering in medicine (2020): 0954411920947850.
		\bibitem{NFL} Sanchez, Erin J., et al. "A reanalysis of football impact reconstructions for head kinematics and finite element modeling." Clinical biomechanics 64 (2019): 82-89.
		\bibitem{NHTSA} National Highway Traffic Safety Administration. “Data.” NHTSA, 18 May 2020, www.nhtsa.gov/data.
		\bibitem{Kleiven2006}Kleiven, Svein. "Evaluation of head injury criteria using a finite element model validated against experiments on localized brain motion, intracerebral acceleration, and intracranial pressure." International Journal of Crashworthiness 11.1 (2006): 65-79.
		\bibitem{zhou2018}Zhou, Zhou, et al. "A reanalysis of experimental brain strain data: implication for finite element head model validation." SAE 62nd Stapp Car Crash Conference, STAPP 2018; Catamaran Resort Hotel San Diego; United States; 12 November 2018 through 14 November 2018. Vol. 2019. SAE International, 2019.
		\bibitem{Bartsch2020} Bartsch, Adam J., et al. "High Energy Side and Rear American Football Head Impacts Cause Obvious Performance Decrement on Video." Annals of Biomedical Engineering (2020): 1-11.
		\bibitem{UMN} Semmel, Sarah. "Multiple Regression in Industrial Organizational Psychology: Relative Importance and Model Sensitivity." (2018).
		\bibitem{Nathans2012} Nathans, Laura L., Frederick L. Oswald, and Kim Nimon. "Interpreting multiple linear regression: A guidebook of variable importance." Practical Assessment, Research, and Evaluation 17.1 (2012): 9.
		\bibitem{Courville2001} Courville, Troy, and Bruce Thompson. "Use of structure coefficients in published multiple regression articles: $\beta$ is not enough." Educational and Psychological Measurement 61.2 (2001): 229-248.
		\bibitem{CA} Ray‐Mukherjee, Jayanti, et al. "Using commonality analysis in multiple regressions: a tool to decompose regression effects in the face of multicollinearity." Methods in Ecology and Evolution 5.4 (2014): 320-328.
		\bibitem{Budescu1993} Budescu, David V. "Dominance analysis: a new approach to the problem of relative importance of predictors in multiple regression." Psychological bulletin 114.3 (1993): 542.
		\bibitem{Prunier} Prunier, Jérôme G., et al. "Multicollinearity in spatial genetics: separating the wheat from the chaff using commonality analyses." Molecular ecology 24.2 (2015): 263-283.
		\bibitem{bootstrapping} Efron, Bradley, and Robert J. Tibshirani. An introduction to the bootstrap. CRC press, 1994.
		\bibitem{CP} Rowson, Steven, and Stefan M. Duma. "Brain injury prediction: assessing the combined probability of concussion using linear and rotational head acceleration." Annals of biomedical engineering 41.5 (2013): 873-882.
		\bibitem{PRHIC} Kimpara, Hideyuki, et al. "Head injury prediction methods based on 6 degree of freedom head acceleration measurements during impact." International Journal of Automotive Engineering 2.2 (2011): 13-19.
		\bibitem{BRIC} Takhounts, Erik G., et al. "Kinematic rotational brain injury criterion (BRIC)." Proceedings of the 22nd enhanced safety of vehicles conference. Paper. No. 11-0263. 2011.
		\bibitem{RVCI} Yanaoka, Toshiyuki, Yasuhiro Dokko, and Yukou Takahashi. Investigation on an injury criterion related to traumatic brain injury primarily induced by head rotation. No. 2015-01-1439. SAE Technical Paper, 2015. 
		\bibitem{VIF} James, Gareth, et al. An introduction to statistical learning. Vol. 112. New York: springer, 2013.
		\bibitem{Hossein Helmet} Vahid Alizadeh, Hossein, Michael G. Fanton, August G. Domel, Gerald Grant, and David Camarillo. "A Computational Study of Liquid Shock Absorption for Prevention of Traumatic Brain Injury." Journal of Biomechanical Engineering (2020)
		\bibitem{Zhao19} Zhao, Wei, and Songbai Ji. "White matter anisotropy for impact simulation and response sampling in traumatic brain injury." Journal of neurotrauma 36.2 (2019): 250-263.
		\bibitem{Li20} Li, Xiaogai, Zhou Zhou, and Svein Kleiven. "An anatomically detailed and personalizable head injury model: Significance of brain and white matter tract morphological variability on strain." Biomechanics and modeling in mechanobiology 20.2 (2021): 403-431.
		\bibitem{Fahlsted21} Fahlstedt, M. et al. Ranking and Rating Bicycle Helmet Safety Performance in Oblique Impacts Using Eight Different Brain Injury Models. Ann Biomed Eng 49, 1097-1109 (2021).
		\bibitem{TimeWindow} Liu, Yuzhe, et al. "Time Window of Head Impact Kinematics Measurement for Calculation of Brain Strain and Strain Rate in American Football." arXiv preprint arXiv:2102.05728 (2021).
		\bibitem{Zhou21} Zhou, Z., Li, X., Liu, Y., Fahlstedt, M., Georgiadis, M., Zhan, X., Raymond, S.J., Grant, G., Kleiven, S., Camarillo, D., 2021. Towards a comprehensive delineation of white matter tract-related deformation. bioRxiv.
	\end{thebibliography}

\end{document}